# First Principle Study on Lead-Free $CH_3NH_3GeI_3$ and $CH_3NH_3GeBr_3$ Perovskite Using FHI-aims Code

H. Abdulsalam[1] and G. Babaji[2]
[1]Department of Physics, Yobe State University, Damaturu P.M.B.1144, Yobe Nigeria.
[2]Department of Physics, Bayero University, Kano P.M.B. 3011, Kano Nigeria.

**Abstract**: An ab-initio calculation in the framework of DFT, as implemented in the FHI-aims package within GGA with the pbe parameterization was carried out in this work. Although methyl ammonium lead iodide, ($CH_3NH_3PbI_3$) has proven to be an effective photovoltaic material, there remains a main concern about the toxicity of lead. An investigation into the possible replacement of $CH_3NH_3PbI_3$ with $CH_3NH_3GeI_3$ and $CH_3NH_3GeBr_3$ as the active layer in perovskite solar cell was carried out. The electronic band structure, bandgap energy and dielectric constants was calculated for $CH_3NH_3GeI_3$ and $CH_3NH_3GeBr_3$. The effect of temperature on linear thermal expansion coefficient, and temperature dependent of lattice constant was studied in the temperature range of 273 to 318 K, also band gap shift due to lattice expansion was studied. The dielectric constants of these materials was also determined. The energy band gap calculated for $CH_3NH_3GeI_3$ and $CH_3NH_3GeBr_3$ at their respective equilibrium lattice constant are 1.606 and 1.925eV respectively.

**Corresponding Author's Email: hadulsalam@ysu.edu.g**

1. Introduction

The research into new types of solar cells is driven by the need for cheaper, clean and sustainable energy. One promising route for departing from the traditional solar cells is the dye-sensitized solar cell (DSSC) (Gratzel, 2009). In this type of cell, the light is harvested by a sensitizer, which may be a dye molecule or a semiconductor quantum dot attached to a wide-band-gap semiconductor of Nano crystalline morphology, typically $TiO_2$ [1]. Dye-sensitized solar cells have stood out among various photovoltaic devices owing to their low cost, simple fabrication procedure, environmental friendliness, and relatively high efficiency. Consequently, DSSCs are promising candidates for the next generation solar cells[2]. The research into new types of solar harvesters for solar cells is driven by the need increase their efficiency and make them more reliable. One promising material for replacing the dye molecule solar harvesters is the organic–inorganic hybrid perovskite. One of the backdrops of dye-molecule harvesters is the fact it exists in liquid form and can dry up unlike the organic-inorganic hybrid perovskites which are in solid form. The introduction of organic–inorganic hybrid perovskites materials as light harvesters and charge carrier transporters; has improve the solar conversion efficiency of DSSCs; such DSSCs are called Perovskites Solar cells (PSC) or Solid-State Dye Sensitized Solar cells (SSDSC). Typically, the light-harvesting active layer is a hybrid organic-inorganic lead or tin halide-base material, the popular among is being methyl ammonium lead iodide, $CH_3NH_3PbI_3$[3]. Perovskite materials such as the methyl ammonium lead halides are cheap to produce and simple to manufacture. Solar cell efficiencies of devices using these materials have increased from 3.8% in 2009[4] to a certified 20.1% in 2014, making this the fastest-advancing solar technology[3]. According to detailed balance analysis, the efficiency limit of perovskite solar cells is about 31%, which approaches the Shockley-Queisser of gallium arsenide which is 33% [5].Their high efficiencies and low production costs make perovskite solar cells a commercially attractive option. Although methyl ammonium lead iodide, ($CH_3NH_3PbI_3$) has proven to be an effective active layer material, there remains a main concern about the toxicity of lead. Lead-based perovskites are a major issue that may prejudice implementation of any PSC technology, both regulation and common sense suggest that PSCs have to become lead free to deliver a sustainable technology[6]. Given the above developments, the determination of a lead free halide perovskite is of outstanding interest.

To design an efficient solar cell device, a deep understanding of underlying material's properties such as chemical composition, mechanical, electrical, and optical properties are required. Quantum mechanical approaches provide a deep understanding of properties of many body systems such as chemical composition, mechanical, electrical, and optical properties. Among the large panel of available theoretical approaches, the density-functional theory (DFT) has become overwhelmingly popular. Its success greatly relies on the existence of efficient computer numerical codes. In these numerical codes the input parameters can be adjusted. The overall principles of DFT are based on the Hohenberg–Kohn's theorems[7]. DFT has a strong versatility especially in the description of the ground state properties of semiconductors and metals. Increase in computing power has afforded further capabilities in system's size that DFT methods can handle.

In this work, an investigation into the possible replacement of $CH_3NH_3PbI_3$ with$CH_3NH_3GeI_3$ and $CH_3NH_3GeBr_3$as the active layer in perovskite solar cell. The band structure analysis and energy bandgap calculation was performed. An investigation into the effect of temperature on lattice constant (lattice expansion) and the effect of lattice constant (lattice expansion) on energy bandgap of the materials was carried out. Furthermore, the dielectric constant was determined.

## 2 Theoretical Background

### 2.1 Density Functional Theory

Density Functional Theory (DFT) is a quantum mechanical technique used in Physics and chemistry to investigate the structural and electronic properties of many body systems. DFT has proved to be highly successful in describing structural and electronic properties in a vast class of materials, ranging from atoms and molecules to simple crystals and complex extended systems (including gasses and liquids). Furthermore DFT is computationally very simple. For these reasons DFT has become a common tool in first-principles calculations aimed at describing or even predicting properties of molecular and condensed matter systems[8].

Traditional methods in electronic structure theory, in particular Hartree-Fock theory and its descendants are based on the complicated many-electron wave function. The main objective of density functional theory is to replace the many-body electronic wave function with the electronic density as the basis quantity. Whereas the many-body wave function is dependent on 3N variables, three special variables for each of the N electrons, the density is only a function of three variables and is a simpler quantity to deal with both conceptually and practically. Consider a system of N interacting (spinless) electrons under an external potential $V(r)$ (usually the Coulomb potential of the nuclei). If the system has a nondegenerate ground state, it is obvious that there is only one ground-state charge density $n(r)$ that corresponds to a given $V(r)$. In 1964 Hohenberg and Kohn demonstrated the opposite, far less obvious result: there is only one external potential $V(r)$ that yields a given ground-state charge density $n(r)$. The demonstration is very simple and uses *a reduction ad absurdum* argument[9].

For many-electron Hamiltonian with ground state Wave function $\psi$ is given by:
$$H = T + U + V \qquad (1)$$
Where $T$ is the kinetic energy, $U$ is the electron-electron interaction, $V$ is the external potential. The charge density $n(r)$ as defined by Hohenberg and Kohn in 1964 is
$$n(r) = N \int |\psi(r_1, r_2 r_3, \ldots r_N)|^2 \, dr_2 \ldots dr_N \qquad (2)$$
Considering a different Hamiltonian: $H' = T' + U' + V'$; with ground state wave function $\Psi'$. Assuming the ground state charge densities are the same; i.e. $n[V] = n'[V']$. Then the following inequality holds:
$$E = \langle \psi' H' \psi' \rangle < \langle \psi H' \psi \rangle = \langle \psi H + V' - V \psi \rangle \qquad (3)$$
That is:
$$E' < E + \int (V(r) - V'(r)) n(r) \, dr \qquad (4)$$
The inequality is strict because $\psi$ and $\psi'$ are different, being eigenstate of different Hamiltonians. By reversing the primed and unprimed quantities, one obtains an absurd result. This demonstrates that no two different potentials can have the same charge density.

The major problem with DFT is the exact functional for exchange and correlation are not known except for the free electron gas. However approximations exist which permits the calculation of certain physical quantities quite accurately. In DFT the most widely used approximation is the local density approximation (LDA), where the functional depends only on the density at coordinate where the functional is evaluated:

$$E_{XC}^{LDA}[n] = \int n(\vec{r})\epsilon_{XC}[n](\vec{r})d\vec{r} \tag{5}$$

Where $\epsilon_{XC}[n]$ is the exchange-correlation energy density of uniform electron gas. In principle the LDA should only work when the density of the electron gas is almost homogeneous. It has been, however, found to give very good results even when the density of electron gas varies rapidly[10].

The generalized gradient approximations (GGA) are still local but also take into account the gradient of the density at the same coordinate, it has the general form:
$$E_{XC}^{GGA}[n] = \int f\big(n(\vec{r}), \nabla n(\vec{r})\big)d^3\vec{r} \tag{6}$$

## 2.2 Electronic Band Structure

The periodic crystal structure is one of the most important aspects of materials science as many properties of materials depend on their crystal structures. One of its most immediate consequences is the arrangement of the electronic states within bands. For semiconductors, many properties are determined from these bands[11]. The electronic band structure of a solid describes those ranges of energy that an electron within the solid may have and ranges of energy that it may not have[12]. Band theory derives these bands and band gaps by examining the allowed quantum mechanical wave functions for an electron in a large, periodic lattice of atoms or molecules. Band theory has been successfully used to explain many physical properties of solids, such as electrical resistivity and optical absorption and forms the foundation of the understanding of all solid-state devices. Band structure calculations take advantage of the periodic nature of a crystal lattice, exploiting its symmetry.

The electronic band gaps of Perovskite materials are determined by the states at the valence band maximum (VBM) and conduction band minimum (CBM). A requisite for PSC to work properly just like in DSSC, the LUMO level of the light harvester should be higher than the conduction band edge (CBE) of anode material[13]; for example in $TiO_2$ which is located at -4.0 eV [14]. This would provide the required driving force for a faster excited state electron injection. The magnitude of the band gap determines the onset of optical absorption and is closely related to the maximum voltage achievable in a photovoltaic device[15]. The energy band gaps of organic-inorganic perovskite increase with increasing lattice parameter, contrary to most general semiconductors like Si and GaAs, this is due to the electronic structure of the Perovskite materials[16]. It was found experimentally that the band gap of $CH_3NH_3PbI_3$ increases with lattice parameter, as evidenced by Photoluminescence (PL) results[17].

## 2.3 Dielectric Constant

The dielectric constant is obtained from the response of the material to an external electric field, it depends of the frequency of the applied electric field and is described by a tensor for anisotropic system. The dielectric tensor $\epsilon_r$ consists of a real part which represents the storage and an imaginary part which represents the loss[18]. The dielectric constant also called relative permittivity is value of the real part of the dielectric tensor at frequency equals to zero[19] i.e. $Re[\epsilon_r(\omega = 0)]$. The amount electric field attenuated in a substance compared to from a vacuum is indicated by its dielectric constant. Dielectric constant determines the magnitude of the coulomb interaction between electron-hole pairs and charge carriers as well as any fixed ionic charges in the lattice, high dielectric constants are required for high efficiency solar cell. Dielectric constant for $CH_3NH_3PbI_3$ is in the range of 5–7[20]

## 2.4 Phonons: Harmonic Vibrations

In reciprocal space, the equation of motion for the vibration of a periodic array of harmonic atoms for each reciprocal vector **q** is determined by dynamical matrix, $D(q)$.

$$D(q)v(q) = \omega^2(q)\, v(q) \quad (7)$$

where $\omega^2(q)$ is eigenvalue, and $v(q)$ is eigenvector of the dynamical matrix $D(q)$ and they completely describe the dynamics of the system in the harmonic approximation, which is a superposition of harmonic oscillators, one for each eigenvalue[21].

The density of state, $g(\omega)$ is an important quantity;

$$g(\omega) = \sum_s \int \frac{dq}{(2\pi)^3} \frac{1}{|\nabla \omega(q)|} \quad (8)$$

It allows the determination of any integrals (e.g. Helmholtz free energy) that only depends on eigenvalue, $\omega$. The Helmholtz (vibrational) free energy $F^{ha}(T,V)$ which has no explicit dependence on the volume $V$ is given by:

$$F^{ha}(T,V) = \int d\omega\, g(\omega) \left( \frac{\hbar\omega}{2} + k_B T \ln\left(1 - e^{\left(-\frac{\hbar\omega}{k_B T}\right)}\right) \right) \quad (9)$$

The heat capacity, $C_V$ at constant volume can be determined from Helmholtz free energy[22];

$$C_V = -T \left( \frac{F^{ha}(T,V)}{\partial T^2} \right)_V \quad (10)$$

## 2.5 Lattice Expansion in the Quasi-Harmonic Approximation

In an ideal harmonic system, which is fully determined by the dynamical matrix $D(q)$, its Hamiltonian does not depend on the volume, this implies that the harmonic Hamiltonian is independent of the lattice parameters, and as a consequence of this, the lattice expansion coefficient $\alpha(T)$ vanishes [21].

$$\alpha(T) = \frac{1}{a}\left(\frac{\partial a}{\partial T}\right)_P \quad (11)$$

To determine the temperature dependence of energy bandgap there is need to determine the lattice expansion. The quasi-harmonic approximation is used to account for the anharmonic effects in the determination of the lattice expansion[23]. The usage of the quasi-harmonic approximation requires the determination of how the phonons, i.e., the vibrational band structures and the associated free energies, change with the volume[22].

## 3.0 Methodology

In this work ab-initio calculation in the framework of density functional theory DFT, as implemented in the FHI-aims package was performed[24]. The geometries of $CH_3NH_3GeI_3$ and $CH_3NH_3GeBr_3$ were built from the structural parameter obtained from existing literature[25] using Visualization for Electronic and Structural Analysis (VESTA)[26] and Avogadro software[27]. The generalized gradient approximation GGA with the blyp parameterization was employed for the evaluation of the exchange-correlation energy. Optimization of the followings configuration parameters; *occupation type (Gaussian or Fermi), charge mix param, initial* *moment, and n max pulay* was also performed. The Gamma-centered grid method has been chosen for sampling the Brillion zone. Full relaxation of the atomic positions within the unit cell was performed following the Broyden-Fletcher Goldfarb-Shanno (BFGS) optimization algorithm. Optimal lattice constant was also determined for each structure. The **k**-paths ($\Gamma - X - M - \Gamma - R - X|M - R$)[28]was used in the band structure analysis. Energy band gap were calculated for the optimized geometries along with the optimized configuration parameters using the FHI-aims code. The next procedures was only performed for those materials that gave a promising energy band gap.

To determine of how the vibrational band structures and the associated free energies, change with the volume of materials. The optimal lattice constant of the materials were determined by finding the minimum of the total energy $E_{DFT}(V)$ and $F^{ha}(T,V)$ by using the Birch-Murnaghan's equation of state[29]. Though, in the canonical ensemble, the relevant thermodynamic potential that needs to be minimized is the free energy $F(T,V)$ which is given by:

$$F(T,V) = E_{DFT}(V) + F^{ha}(T,V) \qquad (12)$$

To account for the volume dependence of $F(T,V)$, the total energy, $E_{DFT}(V)$ and free energy, $F^{ha}(T,V)$ were calculated for a series of lattice constants. Eq. (12) was then evaluated and minimized using the Birch-Murnaghan's Equation of states, the phonopy program package and its FHI-aims interface phonopy-FHI-aims are used here. A FHI-aims python script titled: *Compute_ZPE_and_lattice_expansion.py*[22] is used to perform the above procedure, this script requires the following inputs; optimal (equilibrium) lattice constant, the temperature range (0 to 350$K$) and geometry information. The script gives two output files; one contains temperature, the lattice constant and the lattice expansion coefficient and the other contains the equilibrium lattice constant computed with and without zero point energy (ZPE).and bulk modulus.

The investigation of the temperature dependence of the electronic band gaps of materials was carried out using a second python script titled: *Compute_bandgap_at_different_volumes.py*. Here electronic band structure calculations were performed for geometries constructed using the lattice constants generated from by the first script for temperature range of 273 to 318 $K$. This script gives an output file that contains both the lattice constants and the energy bandgap as a function of temperature. To obtain the linear dielectric tensor in FHI-aims the tag *compute_dielectric* is used. It calculates and output the component of the imaginary and real part of the inter-band and intra-band contribution to the linear dielectric tensor.

## 4.0 Results and Discussion
## 4.1 LUMO, HOMO and Bandgap

The LUMO, HOMO and bandgap calculated is tabulate below along with reported values where available. The LUMO level for $CH_3NH_3GeI_3$ is above the CBE of most anode materials (e.g. $TiO_2$ and $SiO_2$), this indicate that it can serve as good active layers, $CH_3NH_3GeBr_3$ on the other hand show good promise as its LUMO level is just below that of the anode materials. The bandgap of $CH_3NH_3GeI_3$ obtained in this work (i.e.1.606 eV)is close to the reported bandgap range for $CH_3NH_3PbI_3$; 1.662 eV[16] and 1.61 eV[30].The bandgap obtained for $CH_3NH_3GeI_3$ in this work is 4.29 % higher than the reported value of 1.54 eV[31]. Figures 3(a) and3 (b) shows that the band gaps of materials increase with increasing lattice parameter like in most perovskite materials, contrary to most general semiconductors like Si and GaAs[16].

**Table 1** LUMO, HOMO and Energy Band gap

| Material | LUMO (eV) | HOMO (eV) | Band Gap (eV) | |
|---|---|---|---|---|
| | | | This work | Reported |
| $CH_3NH_3GeI_3$ | -3.60616747 | -5.21226324 | 1.606 | 1.54[31]2.0[32] |
| $CH_3NH_3GeBr_3$ | -3.85925193 | -5.78400997 | 1.925 | - |

## 4.2 Lattice Constant

The lattice constants determined for each of the materials are listed in Table 2. There were little available data on lattice parameters of $CH_3NH_3GeI_3$ and $CH_3NH_3GeBr_3$. Although the lattice constant obtained in this work was not compared with other reported values, the lattice constant increases from Br to I., a similar trend was found from Sn and Pb serials[16].

**Table 2** Lattice Constants

| Material | Lattice Constant(Å) | | |
|---|---|---|---|
| | From Single point calculation | From phonopy without ZPE | From phonopy with ZPE |
| $CH_3NH_3GeI_3$ | 5.933 | 5.876 | 5.907 |
| $CH_3NH_3GeBr_3$ | 5.833 | 5.877 | 5.894 |

## 4.3 Linear thermal expansion

Linear thermal expansion coefficient, $\alpha$ given in Eq. (11) was plotted against temperature for each of the materials, as shown in Figures 1(a), in figure 1(b) the same graph was plotted for more clarity for temperature range of 0 to 20 K. The change in lattice constant with temperature is shown in figure 2. Figure 2(a) shows that the linear thermal expansion coefficient does not change constantly with temperature, and it is negative for some very low temperatures, but in Figure 1(b) the negative expansion (contraction) can be seen to occur between the temperature ranges of 0 to 3 K. The temperature range at which the negative expansion occurs is the same for both materials. There is a constant increment of the expansion coefficient at temperatures above $3K$. The negative expansion observed for this materials is similar to those observed in some semiconductors such as Germanium, Silicon, Diamond and Gallium Arsenide[33]. The lattice constant expands with increase in temperature but with contraction at very low temperatures, this is obvious because of the negative linear thermal expansion observed in figure 1(b).

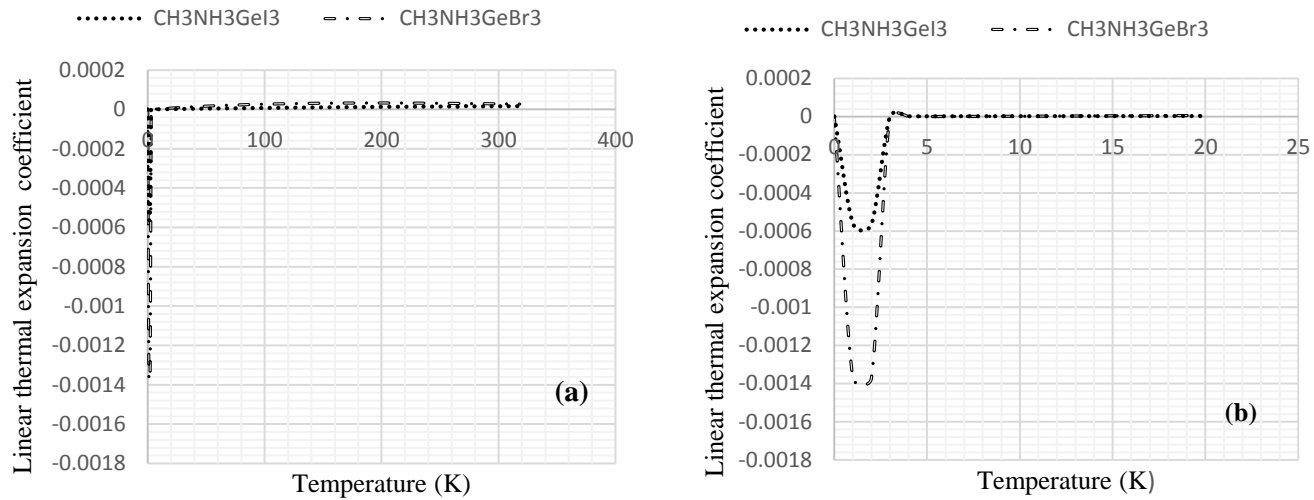

**Fig. 1** Graph of Linear thermal expansion coefficient against temperature for: (a) $CH_3NH_3GeI_3$ (b) $CH_3NH_3GeBr_3$

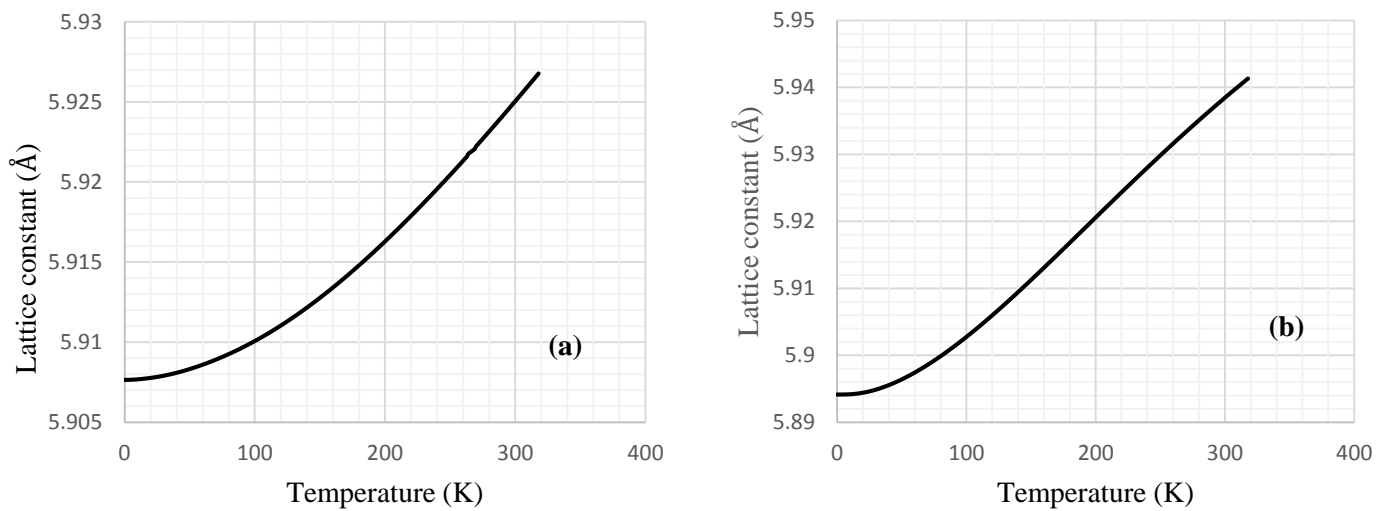

**Fig. 2** Graph of Lattice constant against temperature for: (a) $CH_3NH_3GeI_3$ (b) CH3NH3GeBr3

### 4.4 Real components of dielectric Tensor

The graph of real components of linear dielectric tensor against frequency for cubic directions [100], [010], [001] and [111] are shown in figure 3(a) and 3(b). The dielectric constant (relative permittivity) which is value of the real part of the linear dielectric tensor at frequency equals to zero (i.e. $Re[\epsilon_r(\omega = 0)]$ ) for each of the above direction were given in Table 3. For both materials, the dielectric constant (relative permittivity) at the [100] and [010] directions have very close values, the ones obtained at [001] directions are higher than the ones at [100] and [010] directions. While the relative permittivity at the [111] direction have the highest value. The dielectric constant obtained at [111] direction for $CH_3NH_3GeI_3$(5.79) is close to the reported range of (5-7)[20] for $CH_3NH_3PbI_3$.

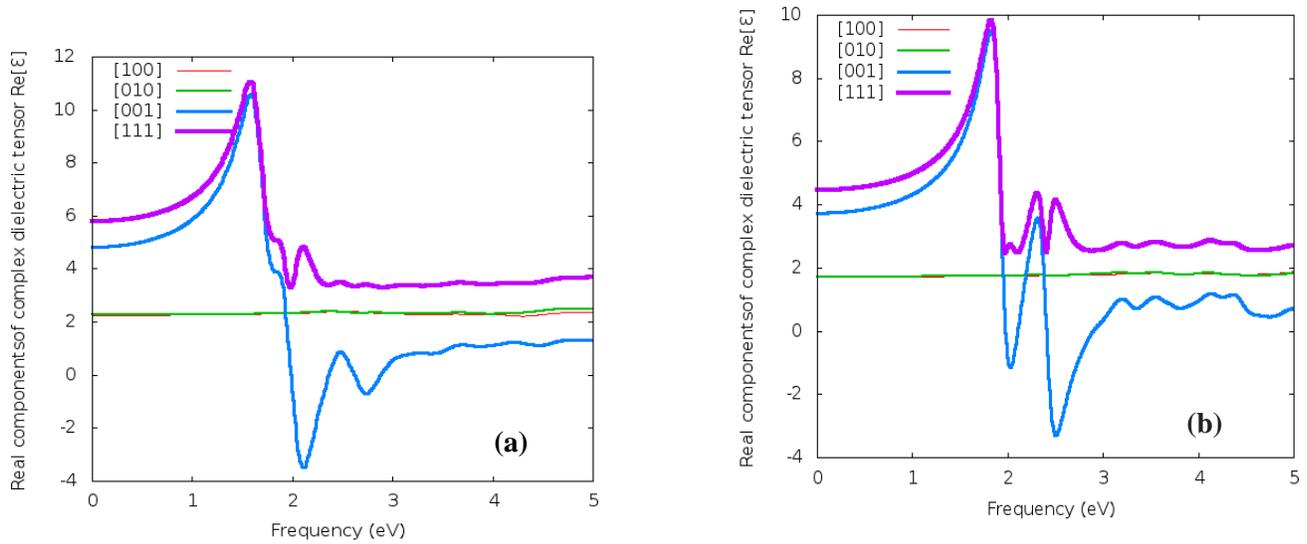

**Fig. 3.** Graph of real components of complex dielectric tensor against frequency for: (a) $CH_3NH_3GeI_3$ (b) $CH_3NH_3GeBr_3$

**Table 3** Dielectric constant ($\epsilon$)

| Material | $Re[\epsilon_r(\omega = 0)]$ | | | |
|---|---|---|---|---|
| | [100] | [010] | [001] | [111] |
| $CH_3NH_3GeI_3$ | 2.26 | 2.27 | 4.83 | 5.79 |
| $CH_3NH_3GeBr_3$ | 1.72 | 1.72 | 3.73 | 4.46 |

## 5.0 Conclusion

All the calculations done in this work were based on Density Functional Theory method as implemented in FHI-aims code. FHI-aims code input parameters were optimized before the actual calculation and estimation were done. The bandgap of $CH_3NH_3GeI_3$ calculated in this work differ from reported result by 4.29 %, the band gaps of materials increase with increasing lattice parameter like in most perovskite materials. The trend observed in the lattice constant of $CH_3NH_3GeI_3$ and $CH_3NH_3GeBr_3$ is similar to that observed in other perovskite materials. Also the phonopy program package and its FHI-aims interface correctly predict the effect of temperature on linear thermal expansion coefficient and lattice constants of the materials, here the temperature dependence observed agrees with other established views.